\documentclass[conference]{IEEEtran}
\IEEEoverridecommandlockouts
\usepackage[%
   paperwidth=210mm,
   paperheight=297mm,
   vmargin={15mm,15mm},
   hmargin={13mm,13mm},
   headsep=12pt,
   footskip=12pt,
   columnsep=18pt
]{geometry}
\usepackage{tikz}
\usetikzlibrary{shapes,arrows,positioning}
\usetikzlibrary{arrows.meta}
\usepackage{cite}
\usepackage{amsmath,amssymb,amsfonts}
\usepackage{algorithmic}
\usepackage{graphicx}
\usepackage{textcomp}
\usepackage{xcolor}
\usepackage{ wasysym }
\def\BibTeX{{\rm B\kern-.05em{\sc i\kern-.025em b}\kern-.08em
    T\kern-.1667em\lower.7ex\hbox{E}\kern-.125emX}}
    
    \usepackage{comment}

\newtheorem{definition}{Definition}
\newtheorem{theorem}{Theorem}
\newcommand{\Sadegh}{\textcolor{red}}
\newcommand{\Vahid}{\textcolor{blue}}
\newcommand{\Ben}{\textcolor{brown}}

\newcommand{\alphabeth}{\Sigma}

\newcommand{\until}{\mathbin{\sf U}}

\newcommand{\nex}{\mathord{\bigcirc}}

\newcommand{\word}{\boldsymbol{\omega}}
\newcommand{\wordt}[1]{{\omega}_{#1}}

\newcommand{\Lab}{\textsf{L}}   
    
\begin{document}


\title{Formal Controller Synthesis for Frequency Regulation Utilising Electric Vehicles}


\author{\IEEEauthorblockN{Ben Wooding}
\IEEEauthorblockA{\textit{School of Computer Science} \\
\textit{Newcastle University}\\
Newcastle-upon-Tyne, UK \\
b.wooding1@newcastle.ac.uk}
\and
\IEEEauthorblockN{Vahid Vahidinasab}
\IEEEauthorblockA{\textit{School of Engineering} \\
\textit{Newcastle University}\\
Newcastle-upon-Tyne, UK \\
vahid.vahidinasab@newcastle.ac.uk}
\and
\IEEEauthorblockN{Sadegh Soudjani}
\IEEEauthorblockA{\textit{School of Computer Science} \\
\textit{Newcastle University}\\
Newcastle-upon-Tyne, UK \\
sadegh.soudjani@newcastle.ac.uk}
}

\maketitle



\begin{abstract}
    In this paper, we propose a formal controller synthesis approach for integrating a population of plug-in electric vehicles in frequency regulation of power systems. This approach is the first application of formal methods to the smart grids in particular to the frequency regulation of power systems. We design and simulate a novel symbolic controller for the Great Britain power system. The proposed controller enhances the frequency response behaviour of the system when encountered with a large outage event. The symbolic controller guarantees the settlement of the after-event's frequency in the specified safe interval and ensures other requirements on the frequency are met.
\end{abstract}

\vspace{10pt}

\section{Introduction}

\tikzstyle{block} = [draw, rectangle, 
    minimum height=3em, minimum width=2em]
\tikzstyle{sum} = [draw, circle, node distance=2.6cm]
\tikzstyle{input} = [coordinate]
\tikzstyle{output} = [coordinate]
\tikzstyle{pinstyle} = [pin edge={to-,thin,black}]

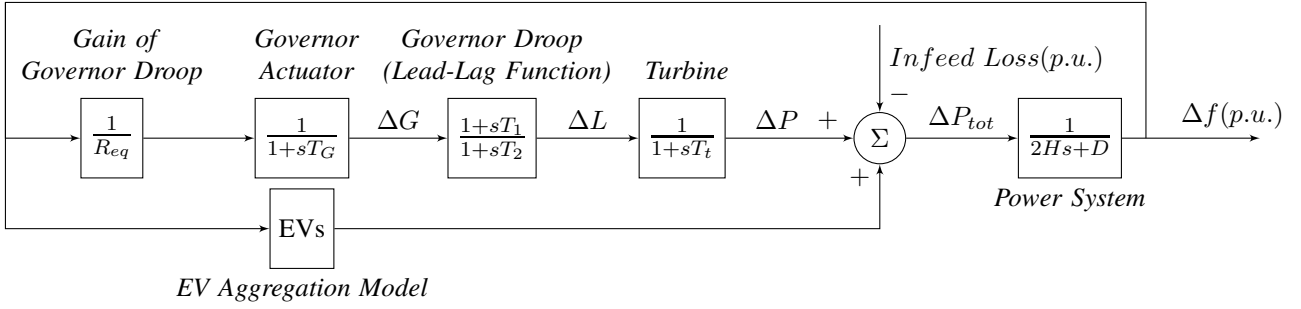
\begin{figure*}[t!]
\centering
\begin{tikzpicture}[auto, node distance = 2.5cm, >=latex', scale = 0.2]
\node[block, label = above:{\begin{tabular}{c} \textit{ Gain of} \\ \textit{Governor Droop} \end{tabular}}](droop){\large \(\frac{1}{R_{eq}}\)};
\node[block, right of = droop, label = above:{\begin{tabular}{c} \textit{ Governor} \\ \textit{ Actuator} \end{tabular}}](GA){\large\(\frac{1}{1 + sT_{G}}\)};
\node[block, right of= GA, label = above:{\begin{tabular}{c} \textit{Governor Droop} \\ \textit{ (Lead-Lag Function)} \end{tabular}}](LL){\large\(\frac{1 + sT_{1}}{1 + sT_{2}}\)};
\node [block, right of = LL, , label = above:{\begin{tabular}{c} \textit{ Turbine} \end{tabular}}] (turbine) {\large\(\frac{1}{1 + sT_{t}}\)};
\node[sum, right of = turbine] (sum) {\(\Sigma\)};
\node[block, right of = sum, label = below:{\textit{Power System}}] (TF) {\large\(\frac{1}{2H_{}s + D}\)};
\node [output, right of=TF] (output) {};
\node[block, below of = GA, node distance = 1.2cm, label = below:{\textit{EV Aggregation Model}}](EV){
EVs
};
\node[input, above of = sum, node distance = 1.5cm](loss){};

\draw[->] (turbine) -- node[pos=0.8]{$+$} node[pos=0.4] {\(\Delta P\)} (sum);
\draw[->] (sum) -- node {\(\Delta P_{tot}\)} (TF);
\draw [->] (TF) -- node [name=f,pos=0.8]{$\Delta f(p.u.)$}(output);
\draw [->] (EV) -| node[pos=0.9] {$+$}(sum);
\draw [->] (GA) -- node {\(\Delta G\)} (LL);
\draw [->] (LL) -- node {\(\Delta L\)} (turbine);
\draw[->] (68,0) |- node [pos = 0.8]{
}(-7,9) |- (droop);
\draw[->](loss) -- node[pos=0.8]{$-$}node[pos = 0.4] {\(Infeed~Loss (p.u.)\)}(sum);
\draw[->] (droop)  -- (GA);
\draw[->] (-7,9) |- (EV.west);

\end{tikzpicture}
    \caption{A simplified GB power system model including EVs for primary frequency response services, based on \cite{MU201306}.}
    \label{fig:tikzdiagram}
\end{figure*}

\begin{figure*}[t]
\centering
\begin{tikzpicture}[auto, node distance = 2cm, >=latex', scale = 0.2]
\node[input](f){};
\node[block, label = above:{\textit{EV Aggregation Model}}](AM){\begin{tikzpicture}
\node[block, right of = f, label = above:{\begin{tabular}{c} \textit{Dead} \\ \textit{Band} \end{tabular}}](deadband){\begin{tikzpicture}
\coordinate (A) at (-.4,-.4);
\coordinate (B) at (-.2,0);
\coordinate (C) at (+.2,0);
\coordinate (D) at (+.4,+.4);
\draw [black, thick] (A)--(B)--(C)--(D);
\end{tikzpicture}};
\node[block, right of = deadband, label = above:{\begin{tabular}{c} \textit{Droop} \end{tabular}}](droop){\large$\frac{1}{R_{ev}}$};
\node[block, right of = droop, label = above:{\begin{tabular}{c} \textit{Transfer} \\ \textit{Function} \end{tabular}}](TF){\large $\frac{1}{sT_{ev} + D}$};
\node[block, right of = TF, node distance = 3cm, label = above:{\begin{tabular}{c} \textit{Saturation}\end{tabular}}](saturation){\begin{tikzpicture}
\coordinate (A) at (-.5,-.4);
\coordinate (B) at (-.3,-.4);
\coordinate (C) at (+.3,+.4);
\coordinate (D) at (+.5,+.4);
\draw [black, thick] (A)--(B)--(C)--(D);
\end{tikzpicture}};
\node[block, right of = saturation,label = above:{\begin{tabular}{c} \textit{Aggregate EV} \\ \textit{Power} \end{tabular}}](k){\large \(p_{av}\)};
\node[block, right of = k, label = above:{\begin{tabular}{c} \textit{Total} \\ \textit{EVs} \end{tabular}}](n){\large \(N_{ev}\)};
\node[input, above of = sum, node distance = 1.5cm](loss){};
\draw[->](deadband) -- (droop);
\draw[->](droop) -- (TF);
\draw[->](TF) -- (saturation);
\draw[->](saturation) -- node{$\Delta u$} (k);
\draw[->](k) -- (n);
\end{tikzpicture}};
\draw[->](-35,-3) -- node[pos=0.3]{ $\Delta f$} (-30,-3);
\draw[->](29.75,-3) -- (35,-3);
\end{tikzpicture}
    \caption{Block diagram of EV frequency response services when power loss occurs, adapted from \cite{MUCHAOXU2018,Izadkhast2015}.}
    \label{fig:EVtikz}
\end{figure*}
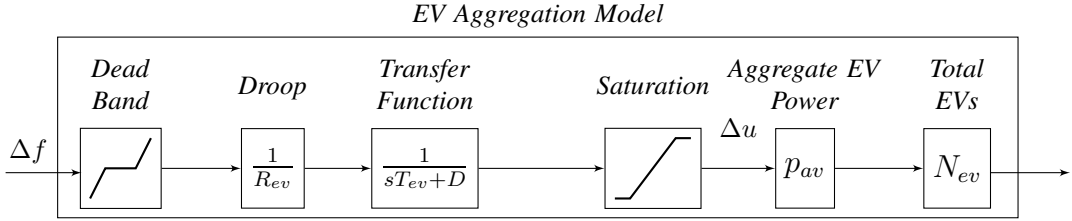

On \(9^{th}\) August 2019, there was a power outage event started in Cambridgeshire, Great Britain (GB), due to a lightning strike that hit overhead transmission lines, affecting about 1 GW demand (i.e. around 5\% of total electric demand). This caused countrywide losses comprised of a 740 MW power station, a 1,200 MW wind farm (on the day outputting 800 MW) and various embedded generation unit losses leading to a total generation loss as part of the initial event of over 2000 MW. The frequency of the system fell to 48.8 Hz, below the statutory limit of 49.5 Hz, at which the automatic protection system known as Low Frequency Demand Disconnection Scheme (LFDD) are triggered to protect the other 95\% demand. Due to the LFDD, over one million customers were affected by the disruption \cite{EnergyEmergenciesExecutiveCommittee2020}. A normal frequency range was restored within 5 minutes but essential services such as transport, health and water were still affected up to two days later. This research paper is inspired by such event.

Frequency response is the reaction to a change in grid frequency, most frequency response occurs from the supply-side where the turning on and off of turbines balance the generation-consumption relationship within the power system. Future smart grid technologies look to use demand-side resources to regulate grid frequency which saves costs, energy and time when power disruption events occur \cite{CALLOWAY2011,TAN201403}. This paper studies primary frequency response of the GB grid based on the model of \cite{MU201306} and shows how a formal controller for plug-in electric vehicles (EVs) could aid frequency recovery during system contingencies.

EVs have been proposed as a means of frequency regulation due to the fast response the EVs can provide to a power disruption event. Reference \cite{ISLAM201911} discusses EVs as essential to the future of smart grids due to gas and diesel vehicles slowly being phased out. Frequency regulation is the most beneficial ancillary service that EVs can provide due to minimal impacts on battery degradation. One study even argues that EV battery life can be extended if EVs take part in demand-side frequency response, compared with regular EV use \cite{UDDIN2017710}. Other EV ancillary services are discussed in \cite{KEMPTON2005268}. EVs respond to frequency events depending on the type of plug-in charger that is being used. Unidirectional chargers receive power from the grid 
and when signalled they stop charging to reduce grid the demand. Bidirectional chargers have the option to discharge energy stored in a EV back into the grid
which leads to wider frequency response services. Bidirectional charging is likely to only be viable for level 2 type chargers, while unidirectional charging would be valid for all other levels 
of charging speed \cite{Yilmaz2013}. Charging strategies are discussed in \cite{Bao2012}. This paper uses a simple model based on \cite{MUCHAOXU2018} and \cite{Izadkhast2015} to simulate the aggregate behaviour of a collection of EVs.

\emph{Formal methods} can be used to achieve frequency response services in the smart grid.
Formal methods give guarantees for safe operation in many safety critical systems. Similarly, \emph{formal verification} is a technique used to verify if systems meet a desired specification. There is a large body of literature in Computer Science for encoding a desired specification in a certain formal language including \emph{linear temporal logic} (LTL) and \emph{computational tree logic} (CTL) \cite{BK08}. Such specifications are able to accurately capture the behaviour of a system over time \cite{Alur2000}.

\emph{Formal synthesis} consists of designing controllers such that the system satisfies a desired specifications, e.g. the states remain in the safe region or reach a target region.
Due to the continuous nature of the state space, \emph{abstraction techniques} are a key component of formal synthesis of systems and a growing area of research. 
A system can be abstracted by partitioning the state space and representing partitions by single points in the abstract state space. 
The mathematical properties of the abstract system can be used to ensure satisfaction of properties in the original system \cite{Girard2007pappas,Alur2000}. From abstraction, LTL properties can be preserved with language equivalent relations, and CTL properties with bisimulations \cite{Alur2000,Mallik2019}. Safety, avoiding "bad" states, and reachability, converging to a winning region, are common requirements of formal specifications. Safety properties can be verified and enforced with control barrier functions  \cite{Ames_2019,lindemann2020control}, and in \cite{girard2005}, zonotopes are used for reachability.

For systems that are affected by uncertainty, Markov processes are used as a modelling framework and probabilistic analysis are employed for computing the likelihood of satisfying a specification. The synthesis of controllers for Markov processes is discussed in \cite{HS_TAC19,SSoudjani,majumdar2019symbolic}.
Data-driven techniques are recently developed for controller synthesis of these class of models \cite{lavaei2020formal,kazemi2020formal}. Available tools for formal verification and synthesis include, but are not limited to SCOTS \cite{Rungger2016}, CORA \cite{Kochdumper2020}, Pessoa \cite{Roy2011} and SpaceEx \cite{Frehse2011} for non-probabilistic systems, and FAUST \cite{FAUST15}, StocHy \cite{CauchiTACAS19}, PRISM \cite{PRISM}, ProbReach \cite{ProbReach}, and AMYTISS \cite{AMYTISS} for probabilisitc systems.

Examples of applying formal methods to smart grids in the literature include a symbolic controller design for time-varying DC Microgrids \cite{Zonetti2019}. The works \cite{SA14TCL} and \cite{MESG13} propose formal modelling and synthesis approaches for demand response of thermostatically controlled loads. Reference \cite{Hackenberg2012} discusses formal software engineering having potential to be applied to the smart grid domain and gives the example of a refrigerator with respect to active power. Formal software engineering techniques are used for self-healing smart grids in \cite{Koutsoumpas2013} and in \cite{Akram2018} the smart grid components are formally described in the Z formal specification language. Formal techniques for smart grid power line communication are discussed in \cite{Uddin2019Riaz}. Unfortunately, all previous results on frequency regulation rely on simulations and do not provide any formal guarantee on ensuring the desired behaviour of the frequency over time. The main contribution of this paper is to use formal synthesis techniques from Computer Sciences to find a controller that incorporates the contribution from EVs in the GB power network and guarantees that the frequency of the network abides by the requirements when a power loss incident happens. 

A simplified model of the GB power system is given in \cite{MU201306}. Aggregate models of EVs are described in \cite{Izadkhast2015,MUCHAOXU2018} in the form of differential equations with nonlinear components. In this paper we adapt these aggregate models of a collection of EVs to generate a baseline controller of our system which we can use for comparison. We also express the requirements on the frequency (always stay in a safe interval, and does not go outside of a smaller interval for more than a specific time period) as a temporal logic formula \cite{BK08}. We then use available software tool SCOTS \cite{Rungger2016} to synthesise a controller for the network that guarantees satisfaction of the temporal formula. SCOTS is a software tool for automatic controller synthesis through discrete abstractions. Linear and non-linear differential equations are over-approximated with finite-state symbolic models and controllers are obtained in the form of finite-state machines \cite{Rungger2016}. The closed-loop symbolic model satisfies the specification, thus the original system also satisfies the specification due to the symbolic approximation including all the behaviours of the original system.

In brief, the novel aspects of this work are summarized as follows:
\begin{itemize}
    \item A formal controller synthesis approach for integrating a population of EVs in the power system;
    \item Application of formal methods in frequency regulation of the network;
    \item Design and simulation of a novel symbolic controller for the GB power system;
    \item The proposed controller, enhances the frequency response behaviour of the system when encountered with a large outage event;
    \item The symbolic controller guarantees the settlement of the after-event's frequency in the specified safe interval.
\end{itemize}

This paper is organised as follows. Section~\ref{sec:freq_control} provides the current requirements on frequency of the grid in case of power loss. We also provide in this section the GB model and the baseline controller adapted from the literature for integration of EVs. Section~\ref{sec:Temporal_logic} shows how we can write down the requirements on the frequency as a temporal logic formula. Section~\ref{sec:Formal_Synthesis} provide our formal synthesis approach for finding a controller with guarantees on satisfaction of the requirements. In section~\ref{sec:Results} we present the simulations of our controller comparing it to the baseline controller. Finally in section~\ref{sec:Conclusion} we conclude on all our findings.


\section{Frequency Control}
\label{sec:freq_control}
The system described below 
is a relevant representation of the frequency control in the GB system and can be used to develop a control logic.
\subsection{Frequency Regulation}
\label{Frequency Regulation}

Frequency is defined as the number of alternating current cycles per second (in Hertz) at which a system is running \cite{TheGridCode2020}. The Electric System Operator (ESO) increases or decreases system frequency using reserve and response services \cite{NationalGrid2016}. Positive service response increases generation or reduces demand while negative service response completes the opposite. Positive response services provide power within seconds and are automatically triggered by local frequency readings while reserve services provide power after an instruction is received. If demand is greater than generation the frequency goes down,  if demand is less than generation the frequency increases. The declared frequency of the GB grid is 50 Hz \cite{statutoryinstrumentsno2665}.

\subsection{Requirements on Frequency}

The focus of this paper will be on events of infrequent infeed losses of 2000 MW, similar to the sequence of events mentioned in \cite{EnergyEmergenciesExecutiveCommittee2020} that caused a 2000 MW total loss within a short period of time.
 When such large losses occur protocols such as LFDD are triggered to return stability to the system \cite{TheGridCode2020}. The current accepted maximum \emph{normal infeed loss} for the GB grid is 1320 MW, while the maximum \emph{infrequent infeed loss} is 1800 MW \cite{NETS2019}. A containment zone is given for -0.8 Hz, this value is the maximum frequency deviation allowed for a loss greater than the normal infeed loss. For a normal infeed loss, the maximum deviation should stay within the statutory limits of 50 \(\pm\) 0.5 Hz \cite{Smith2016}. For plants taking part in frequency regulation, a droop characteristic of 3-5\% is expected \cite{TheGridCode2020}. Frequency conditions are required to have a steady state within statutory limits for normal infeed losses and in the case of infrequent infeed loss, a violation should occur for no more than 60 seconds \cite{NETS2019}.
 
 \subsection{The GB Model \label{GB_Model}}
Figure~\ref{fig:tikzdiagram} shows the GB grid model used in this paper that consists of responsive synchronous plants and an aggregate group of EVs. The synchronous plants model is discussed in depth in \cite{MU201306}. Our paper is desired to be a proof of concept not a replication of any specific event therefore we set the values using table~\ref{table:1}. An extensive discussion on modelling of conventional power plants can be found in \cite{kundur1994}. We have included an aggregate model of EVs that are in charging mode and if included in frequency response services, they will contribute to primary frequency control. There are three main frequency response conditions being considered for the EVs, depending on the charging strategy:
\begin{enumerate}
    \item EVs do not participate in frequency response services and continue to charge;
    \item EVs participate in primary frequency response when plugged in using unidirectional chargers;
    \item EVs participate in primary frequency response when plugged in using bidirectional chargers.
\end{enumerate}

\subsection{Baseline Controller}
\label{Baseline Controller}

The adapted aggregate model for these EVs is shown in Figure \ref{fig:EVtikz}. The participation factor is the proportion of vehicles available to contribute to frequency control. In the baseline controller this is calculated by the components between (and including) the dead-band and saturation. The saturation in the system is used to determine the participation giving a value between 0 and 1. The participation value is multiplied by the power per unit (\(p_{av}\)) and the number of vehicles (\(N_{ev}\)) to form the total power provided for frequency regulation. For bidirectional charging, energy can be discharged back into the grid if necessary so twice as much power is available per vehicle. It does not include moving vehicles as these would not contribute to charging demand in the system \cite{Izadkhast2015}.

For this paper, participation will be considered the input of the model. The output of the system is the frequency with respect to the time. As the frequency deviates from the nominal value the controller decides to increase the participation of the EVs in the system to provide response services to return the frequency back to a steady state as near to the nominal value as possible. We take the baseline controller which is adapted from \cite{MUCHAOXU2018,Izadkhast2015} and compare it with our formal synthesis approach that finds a symbolic controller with respect to the requirements on the frequency.


\begin{figure}[t]
\includegraphics[width=\columnwidth]{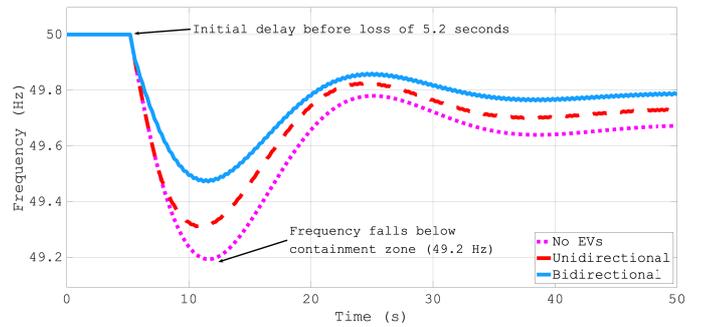}
\caption{Frequency control under 2000 MW infrequent infeed loss using different EV charging strategies.}
\label{fig:2000MWLoss}
\end{figure}

\begin{table}[t]
\caption{values used for simulation adapted from \cite{MU201306,Izadkhast2015,MUCHAOXU2018}.}
\label{table:1}
\centering
\begin{tabular}{ l|c|c}
 \hline
 \vspace{1pt}
 \textbf{Parameters} & \textbf{Unidirectional Value} & \textbf{Bidirectional Value}\\
 \hline
  \vspace{1pt}
 1/$R_{eq}$ & $-5$ & $-5$\\ 
 $T_G$ & $2.5$ & $2.5$\\
 $T_t$ & $0.5$& $0.5$\\
 $T_1$ & $2$& $2$\\
 $T_2$ & $12$ & $12$\\
 $D$ & $1.0$ & $1.0$\\
 $H$ & $4$ & $4$ \\
 $T_{ev}$ & $0.035$ & $0.035$\\
 $R_{ev}$ & $0.5$ & $0.5$\\
 $p_{av}$ & $0.028$ & $0.056$\\
 $N_{ev}$ & $25,000$ & $25,000$\\
 deadband & $50 \pm 0.15$ & $50 \pm 0.15$\\
 \hline
\end{tabular}
\end{table}

\subsection{Baseline Simulation}
When plotting the values from sections \ref{GB_Model} and \ref{Baseline Controller}, Figure~\ref{fig:2000MWLoss} is given. It can be seen that for losses of 2000 MW with no EV input to frequency regulation the containment limit of 49.2 Hz is breached, the system is in a delicate position and a large effort is required to return the frequency to stable conditions. Introducing EVs to primary frequency regulation when considering unidirectional charging improves the response of the system to large power losses. The frequency falls below the statutory limits, but since we consider an infrequent infeed loss this is acceptable, should the system return to the limits within 60 seconds. In the case of both charging strategies this is true. Introducing bidirectional charging improves the recovery even further.

However, simulation will struggle to fully model a real GB system. Change of system's inertia due to the infeed loss is not considered within the system and a decrease in inertia leads to a larger rate of change of frequency. This means the maximum frequency loss could be greater than simulated and so suggested techniques may not be valid in practice. We therefore use these results as a basis for comparison.

Therefore the contribution of the paper is not the simulation of theoretical results but the design of a controller with a given formal specification for how the system should behave. In this regard the system will be able to show mathematically that a specification holds, using techniques such as over-approximation to provide formal proofs for the system. This paper is a proof of concept with the implications of being extended to more complex and real-time systems.

\section{Temporal Logic}
\label{sec:Temporal_logic}

Temporal logic is a formalism for specifying desired properties of systems that evolve over time. Examples of such behaviours include staying in a safe region of the state space, reaching some target region, visiting some region infinitely often, and so on. \emph{Linear temporal logic} is a logic that provides a high-level language for describing such desired behaviour. Formulas in this logic are constructed inductively by using a set of atomic propositions and combining them via Boolean operators. This logic is primarily employed for the study of temporal behaviour of finite-state systems  \cite{BK08}. Recent research has focused on extending the use of this logic on continuous-space deterministic systems \cite{tabuada09} and stochastic systems \cite{SSoudjani,FAUST15,HS_TAC19}. In this paper, we consider LTL for specifying the desired behaviour of frequency of the grid.

LTL uses Boolean symbols such as disjunction ``$\lor$", negation ``$\lnot$" and conjunction ``$\land$". The symbol ``$\lozenge \psi$" is used to denote that some event $\psi$ will eventually happen at some point in the future. The symbol ``$\square \psi$" signifies that $\psi$ must always be true at all time in the future. The symbol ``$\ocircle \psi$" states that $\psi$ must hold in the next time instant. Similarly, ``$\psi_1 \until \psi_2$" is true if $\psi_1$ is true continuously until a time step in the future where $\psi_2$ is true. A detailed precise definition of LTL is provided in the appendix.

\subsection{Formalising the Specification for Frequency}
As described in Section~\ref{Frequency Regulation}, the acceptable behaviours of the frequency as a function of time when considering power loss in the GB grid are given in natural language. We can express these specific behaviours formally in LTL as follows.

First, the frequency should never drop below the containment zone ($C_{zone} = 49.2$ Hz) as at this frequency larger scale frequency response is required to return the system to normal limits and can include load shedding which is hugely disruptive. This can be written as the safety specification
\begin{equation}
\psi_1 := \square (f \ge  C_{zone}).
\end{equation}
Second, the frequency should remain within the statutory limits ($Stat_{lim} = [49.5,50.5]$ Hz)
for any \textit{normal power losses} (i.e., $loss \leq N_{loss}$ with a predefined $N_{loss}$). This can be represented as the LTL formula
\begin{equation}
\psi_2 := (loss \leq N_{loss}) \implies \square (f \in Stat_{lim}).
\end{equation}
Third, for \emph{infrequent infeed losses} specified by the inequality $loss \geq I_{loss}$ with a predefined $I_{loss}$, the frequency must return within $60$ seconds to the statutory limits whenever it leaves that limit. This can be written as the LTL formula
\begin{equation}
\label{eq:Iloss}
\psi_3 := (loss\ge I_{loss}) \implies \lozenge^{60}(f \in Stat_{lim})
\end{equation}
Note that $\lozenge^{60}$ means the condition holds eventually within the next $60$ seconds.
Finally, the desired behaviour of the frequency can be written as
\begin{equation}
\psi = \psi_1\land\psi_2\land\psi_3.
\end{equation}

\medskip
\noindent\textbf{Specification for designing the controller.} As we are interested in an infrequent infeed loss with specification~\eqref{eq:Iloss}, we focus on reachability and show that LTL has the capability of defining a much richer class of behaviours. In particular, we consider a two-stage controller for the frequency regulation. The first controller is responsible for bringing the frequency inside an interval $I_1$ and the second controller is responsible for bringing the frequency inside a smaller interval $I_2\subset I_1$.
\begin{align}
\psi\, := \,\square (f \ge  & C_{zone})\, \land\, \left[\neg(f \in I_1) \implies \lozenge (f \in I_1)\right] \,\land\,\nonumber\\
& \left[(f\in I_1 \land f\not\in I_2)\implies \lozenge (f \in I_2)\right].
\label{eq:spec}
\end{align}
This specification reduces the pressure on the first controller by bringing the frequency inside the smaller interval $I_2$ in multiple phases. Note that since we only consider primary frequency response, it is not necessary for the frequency to return to $50$ Hz as other response schemes would respond in real-time scenarios to aid the full recovery. Therefore, we have not considered any requirement in $\psi$ on the steady state being at $50$ Hz.


\section{Formal Controller Synthesis}
\label{sec:Formal_Synthesis}

In this section, we discuss how to formally design a controller for integrating EVs in the grid such that the frequency satisfies the desired behaviour. Such a formal controller design requires that the time evolution of the system is written down as a dynamical system with differential equations affected by inputs and disturbances.

\subsection{Grid as a Dynamical System}

We can represent the simplified grid model of Fig.~\ref{fig:tikzdiagram} as a dynamical system by converting the transfer functions into differential equations. The dynamics of such a system can be written as
\begin{align}
\dot{f}(t) & = \frac{1}{2H}p(t)+ \frac{p_{av} \times N_{ev}}{2H}u(t) - \frac{loss}{2H} -\frac{D}{2H}f(t)\nonumber\\
\dot{g}(t) & = \frac{1}{T_{g}R_{eq}}f(t) -\frac{1}{T_{g}}g(t)\nonumber\\
\dot{l}(t) & = \frac{T_{1}}{T_{2}T_{g}R_{eq}}f(t)+ \frac{T_{g} - T_{1}}{T_{2}T_{g}}g(t) - \frac{1}{T_{2}}l(t)\nonumber\\
\dot{p}(t) & = \frac{1}{T_{t}}l(t) -\frac{1}{T_{t}}p(t).\label{eq:dynamical_system}
\end{align}
Using equation \eqref{eq:dynamical_system}, a state space model can be constructed of the form
\begin{equation}
    \dot{x}(t) = A x(t) + B u(t) + B_w w(t),
    \label{state_eqn}
\end{equation}
where $x =[f,p,g,l]^{T}\in\mathbb R^4$ is the state vector (shifted around their nominal values), $u\in[0,1]$ is the participation ratio as the input, and $w\in\mathbb R$ is the power loss. The state matrices are
\begin{equation}
\label{eqn:state vectorA}
    A = \begin{bmatrix}
        \frac{-D}{2H}  & 0 & 0 & \frac{1}{2H}\\
        
        \frac{1}{T_{g}R_{eq}} & \frac{-1}{T_{g}} & 0 & 0\\
        
        \frac{T_{1}}{T_{2}T_{g}R_{eq}}  &  \frac{T_{g} - T_{1}}{T_{g}T_{2}} & \frac{-1}{T_{2}} & 0\\
        
        0 & 0  & \frac{1}{T_{t}}  & \frac{-1}{T_{t}}\\
        \end{bmatrix}
\end{equation}
\begin{equation}
\label{eqn:state vectorB}
    B = \begin{bmatrix}
        \frac{p_{av} \times N_{ev}}{2H}\\
        0\\
        0\\
        0
        \end{bmatrix},\quad
         B_w = \begin{bmatrix}
        \frac{-1}{2H}\\
         0\\
        0\\
        0
        \end{bmatrix}
        .
\end{equation}
The power loss is treated as a disturbance $w(t)$ which is bounded by the maximum power loss.

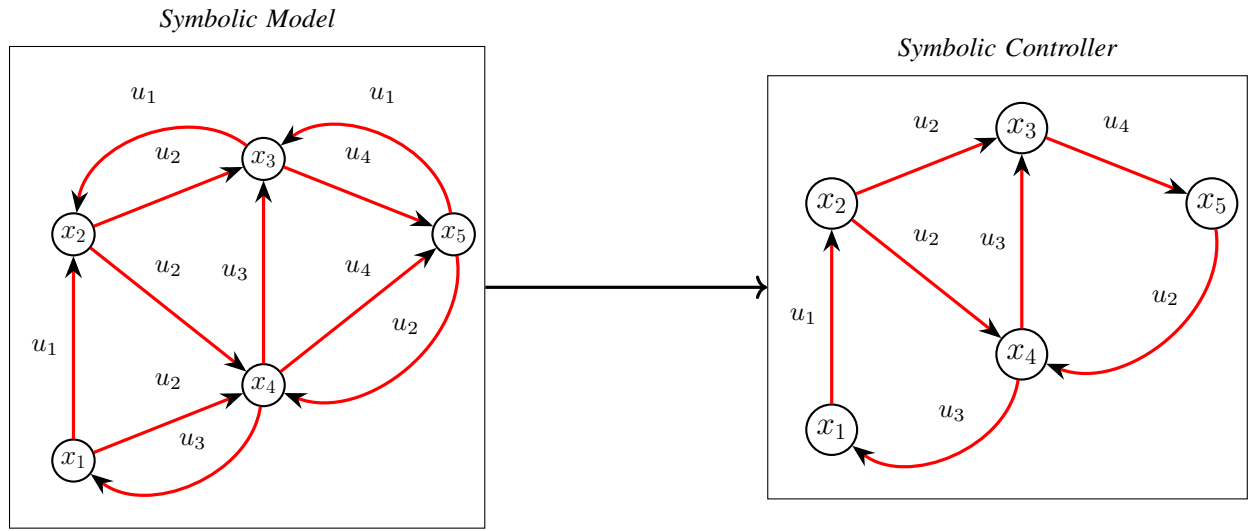
\begin{figure*}[t!]
    \centering
    
\begin{tikzpicture}

\node[block, node distance = 10cm, label = above:{\begin{tabular}{c} \textit{Symbolic Model}\end{tabular}}](sym_model){\begin{tikzpicture}
\begin{scope}[every node/.style={circle,thick,draw, scale = 0.5}]
    \node (A) at (0,0) {\huge $x_1$};
    \node (B) at (0,3) {\huge $x_2$};
    \node (C) at (2.5,4) {\huge $x_3$};
    \node (D) at (2.5,1) {\huge $x_4$};
    \node (E) at (5,3) {\huge $x_5$} ;
\end{scope}

\begin{scope}[>={Stealth[black]},
              every edge/.style={draw=red,very thick}]
    \path [->] (A) edge node[left] {$u_1$} (B);
    \path [->] (B) edge node[above] {$u_2$} (C);
    \path [->] (B) edge node[above] {$u_2$} (D);
    \path [->] (A) edge node[above] {$u_2$} (D);
    \path [->] (D) edge node[above] {$u_4$} (E);
    \path [->] (D) edge node[left] {$u_3$} (C);
    \path [->] (D) edge[bend left = 60,above] node {$u_3$} (A);
    \path [->] (C) edge node[above] {$u_4$} (E);
    \path [->] (E) edge [bend right=60,above] node {$u_1$} (C);
    \path [->] (E) edge [bend left=60,above]node {$u_2$} (D);
    \path [->] (C) edge[bend right=60,above] node {$u_1$} (B); 
\end{scope}
\end{tikzpicture}};

\node[block, right of = sym_model, node distance = 10cm, label = above:{\begin{tabular}{c} \textit{Symbolic Controller}\end{tabular}}](sym_control){\begin{tikzpicture}
\begin{scope}[every node/.style={circle,thick,draw,scale=0.6}]
    \node (A) at (0,0) {\huge $x_1$};
    \node (B) at (0,3) {\huge$x_2$};
    \node (C) at (2.5,4) {\huge$x_3$};
    \node (D) at (2.5,1) {\huge$x_4$};
    \node (E) at (5,3) {\huge$x_5$} ;
\end{scope}

\begin{scope}[>={Stealth[black]},
              every edge/.style={draw=red,very thick,above}]
    \path [->] (B) edge node {$u_2$} (D);
    \path [->] (A) edge node[left] {$u_1$} (B);
    \path [->] (B) edge node {$u_2$} (C);
    \path [->] (D) edge[bend left = 60] node {$u_3$} (A);
    \path [->] (C) edge node {$u_4$} (E);
    \path [->] (D) edge node[left] {$u_3$} (C);
    \path [->] (E) edge [bend left=60]node {$u_2$} (D);
\end{scope}
\end{tikzpicture}};

\path[->,very thick](sym_model) edge (sym_control);
\end{tikzpicture}
\caption{\textbf{Representation of the Symbolic Model $S_a$ and the Symbolic Controller $C_a$.}  The symbolic model shows the state space $X_a$ with \{$x_1,x_2,\ldots, x_5\} \in X_a$ representing partitions of the continuous state space and the input space $U_a$ with values \{$u_1,\ldots,u_4 \} \in U_a$. The transition relation $F_a$ is shown graphically. The symbolic controller stores the appropriate inputs $u_a$ for each state $x_a$. Thus the symbolic controller is treated as a lookup table, providing the input when the system is in a given state to guarantee specification of a satisfaction.}
    \label{fig:my_label}
\end{figure*}{}[t!]


\subsection{Symbolic Model of the Grid}
\begin{definition}
\label{def1}
    A symbolic model of dynamical system \eqref{eq:dynamical_system} for a sampling time $\tau$ is a transition system of the form $S_a := (X_a, U_a, F_a)$, where $X_a$ is a finite partition of the state space of \eqref{eq:dynamical_system}, $U_a$ is a finite subset of input set of \eqref{eq:dynamical_system}, and $F_a:X_a\times U_a\rightarrow 2^{X_a}$ is a transition relation with $2^{X_a}$ being the power set of $X_a$.
\end{definition}
The transition relation $F_a(x_a,u_a)$ is defined as follows:
compute all state trajectories of \eqref{eq:dynamical_system} starting from partition set $x_a$ under the input $u_a$ and for all possible values of the disturbance; then $x'_a\in F_a(x_a,u_a)$ if $x'_a$ intersects with this set of trajectories after a fixed sampling time $\tau$. Fig.~\ref{fig:my_label} (left) shows a graphical representation of a symbolic model.

\begin{theorem}
The particular construction of the symbolic model $S_a$ implies that $S_a$ overapproximates the trajectories of the original model. Thus if a controller is found on the symbolic model that satisfies a given specification, the original model will also satisfy the same specification for all disturbance trajectories. 
\end{theorem}

Available tools for computation of symbolic models and design of symbolic controllers include SCOTS~\cite{Rungger2016} and Pessoa \cite{Roy2011}. These tools usually rely on overapproximating the trajectories of the original model using growth bounds that depend on continuity properties of the differential equations \eqref{eq:dynamical_system}. They also use fixed point computations for finding symbolic controllers. In this paper, we use SCOTS for designing symbolic controllers that has the ability to find controllers for reach, reach-avoid and safety specifications. The safety enforces a specification which guarantees a system is always in a safe state. The reach and reach-avoid focus on eventually reaching a safe state from the current state with the option of avoiding any bad states along that trajectory.
The computations in SCOTS are implemented in C++ language with a MATLAB interface to view the symbolic state space along with simulating the closed loop system~\cite{Rungger2016}.




\subsection{Symbolic Control for the Grid}

A symbolic controller $C_a$ for the symbolic model $S_a := (X_a, U_a, F_a)$ defined in Def.~\ref{def1} is in the form of $u_a = C_a(x_a)$ that assigns any partition set $x_a\in X_a$ to an input $u_a\in U_a$ in order to satisfy the given specification on $S_a$. Such a controller is used to construct a controller $C$ for the original system \eqref{state_eqn} as follows. We have $u(t) = C(x(t))$ with $x(t)\in x_a$ and $u(t) = C_a(x_a)$. In other words, the partition set of $x(t)$ is identified and the input related to that partition set in the symbolic controller is selected as the input for the original system. Fig.~\ref{fig:my_label} (right) shows a graphical representation of a symbolic controller.

In the construction of the symbolic model $S_a$, we select the working region of state variables as $f \in[-1,0.1]$, $p \in[0,3]$, $g \in[0,2]$, and $l \in[0,2]$. The values are generally chosen based on the time constants of the blocks in Fig.~\ref{fig:tikzdiagram} and the range of inputs of these blocks. Adjustments are made to reduce computation time in simulation. Note that these are the states shifted around their nominal values. We partition this working region along each dimension with discretisation $0.05$.
For the input $u \in [0;1]$, discrete steps of $5\%$ of the total input range is considered.
From these partition sets as symbolic states and inputs, a symbolic model and a growth bound are calculated. The growth bound is calculated by taking the Jacobian of the right-hand side of \eqref{state_eqn}  in the form of a Metzler matrix. This is the abstraction of the original system and we compute the transition relation of this new system for our fixed point computations. The fixed point computation of the reach specification using our target range is then calculated giving us a formally synthesised controller. The results of these controllers will be discussed in Section~\ref{sec:Results}.


\section{Implementation Results}
\label{sec:Results}
In this section, we apply our symbolic controller synthesis to the model of the GB power grid and compare it with the baseline controller of Fig.~\ref{fig:EVtikz} proposed in \cite{MUCHAOXU2018,Izadkhast2015}. We have used SCOTS for the design of the symbolic controllers and implemented the simulations in MATLAB on a machine equipped with Intel Core i5-7267U 3.1GHz CPU and 8GB RAM. Computing each of the two controller's reach function takes approximately 28 seconds for unidirectional EVs and 31 seconds for bidirectional EVs.

\subsection{Simulations with a Multi-phase  Controller}

\begin{figure}
\includegraphics[width=\columnwidth]{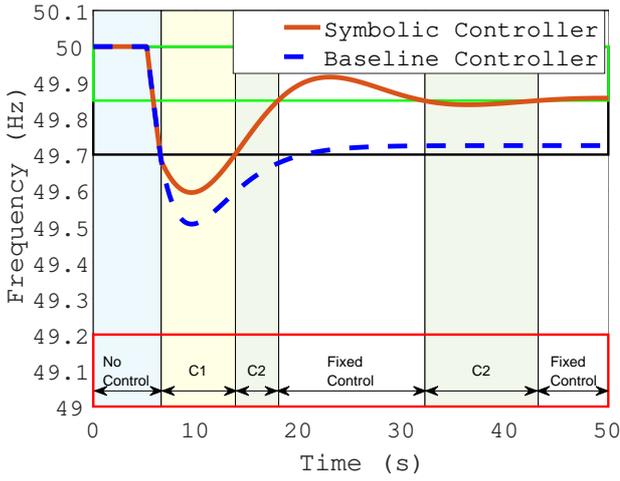}
\caption{Symbolic control for frequency regulation with bidirectionally charged EVs. The frequency under the synthesised symbolic controller satisfies the specification $\psi$ in \eqref{eq:spec} with $C_{zone} = 49.2$, $I_1 = [49.70,50]$ and $I_2 = [49.85,50]$ Hz, but the baseline controller of Fig.~\ref{fig:EVtikz} is unable shape the frequency with respect to~$\psi$.}
\label{fig:bi_freq}
\end{figure}

\begin{figure}
\includegraphics[width=\columnwidth]{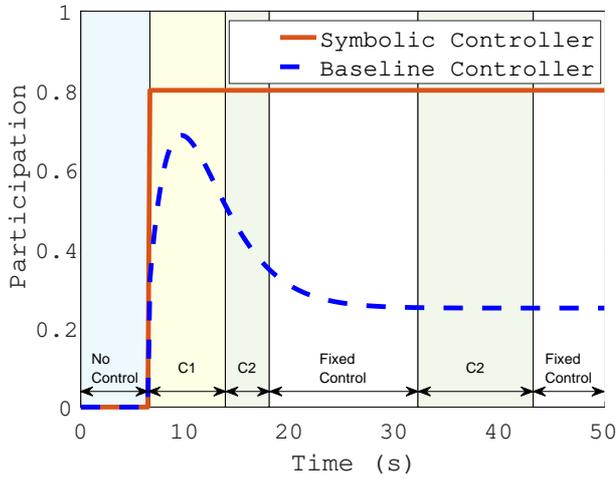}
\caption{Percentage of participation of bidirectionally charged EVs as a function of time obtained from our synthesis approach to satisfy $\psi$ in \eqref{eq:spec} and from the baseline controller of Fig.~\ref{fig:EVtikz}.}
\label{fig:bi_input}
\end{figure}

\begin{figure}
\includegraphics[width=\columnwidth]{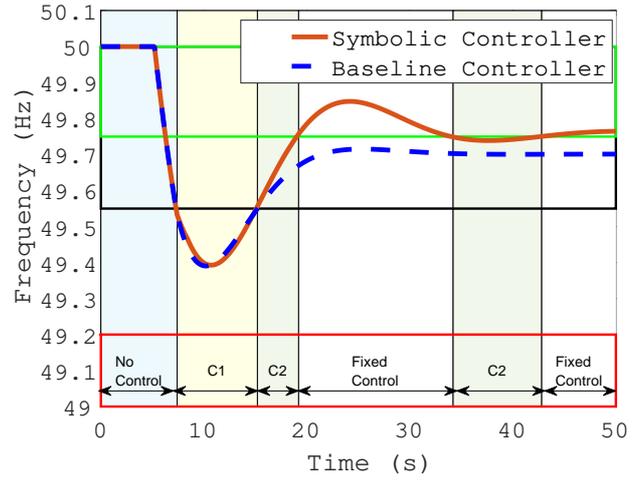}
\caption{
Symbolic control for frequency regulation with unidirectionally charged EVs. The frequency under the synthesised symbolic controller satisfies the specification $\psi$ \eqref{eq:spec} with $C_{zone} = 49.2$, $I_1 = [49.55,50]$ and $I_2 = [49.75,50]$ Hz, but the baseline controller of Fig.~\ref{fig:EVtikz} is unable shape the frequency with respect to~$\psi$.}
\label{fig:uni_freq}
\end{figure}

\begin{figure}
\includegraphics[width=\columnwidth]{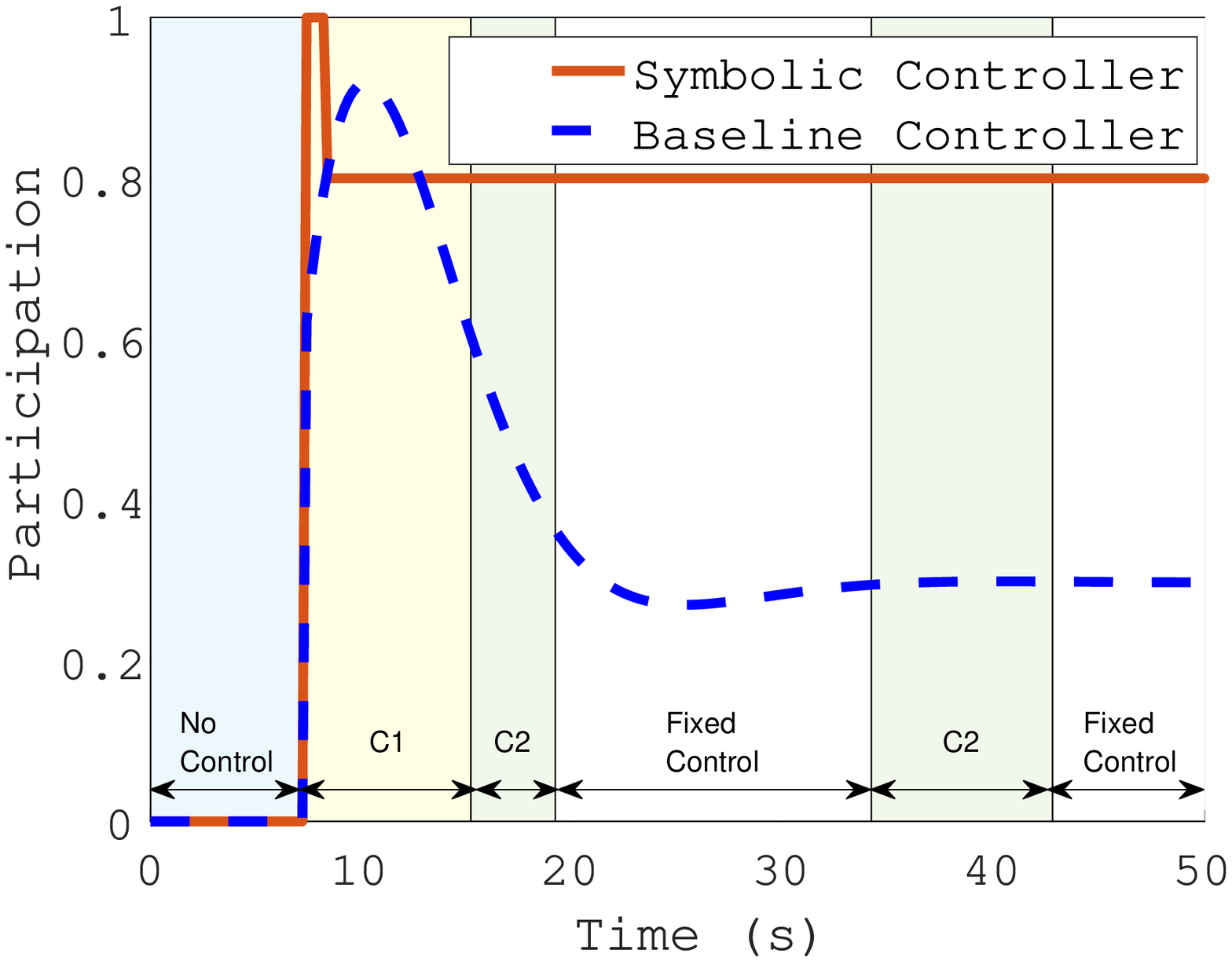}
\caption{Percentage of participation of unidirectionally charged EVs as a function of time obtained from our synthesis approach to satisfy $\psi$ in \eqref{eq:spec} and from the baseline controller of Fig.~\ref{fig:EVtikz}.}
\label{fig:uni_input}
\end{figure}

We have designed a symbolic controller for satisfying the specification $\psi$ in \eqref{eq:spec}. The results of the frequency response are presented in Fig.~\ref{fig:bi_freq} and Fig.~\ref{fig:uni_freq} for respectively bidirectionally and unidirectionally charged EVs.
The containment zone $(f\le C_{zone})$ that should not be visited is shown in these figures with a box having red edges. The target regions $f\in I_2$ and $f\in I_1$ are shown with boxes having respectively green and black edges.
Our symbolic controller uses the following phases control strategy once the power loss occurs:
\begin{enumerate}
    \item If the state is still within the larger target region $I_1$, no participation of EVs is required ($u=0$). This phase is highlighted in light blue in the figures with name ``No Control".
    \item Whenever the frequency leaves the larger target region $I_1$, a low-level symbolic controller is activated to bring the frequency inside $I_1$. This phase is highlighted in light yellow in the figures with name C1.
    \item When the frequency is inside $I_1$ but outside of the smaller target region $I_2$, a second low-level symbolic controller is activated to bring the frequency inside $I_2$, This phase is highlighted in light green in the figures with name C2.
    \item Finally, if the frequency goes inside the smaller target region $I_2$, the last value of participation is used. This phase is in white colour in the figures with name ``Fixed Control".
\end{enumerate}
We have designed symbolic controllers C1 and C2 by solving two reachability problems with target regions $I_1$ and $I_2$ using SCOTS. 
We have selected $I_1 = [49.70,50]$, $I_2 = [49.85,50]$ Hz for bidirectionally charged EVs and $I_1 = [49.55,50]$, $I_2 = [49.75,50]$ Hz for unidirectionally charged EVs. 
The results of
the required participation are presented in Fig.~\ref{fig:bi_input} and Fig.~\ref{fig:uni_input}.

\begin{table}
\caption{Steady state frequency of Baseline Controller for different deadband thresholds (Hz)}
\label{table:2}
\centering
\begin{tabular}{ c|c|c}
 \hline
  \vspace{1pt}
\textbf{Deadband} & \textbf{Unidirectional SS} & \textbf{Bidirectional SS}\\
 \hline
 $50 \pm 0.00$ & 49.73 & 49.77\\ 
 $50 \pm 0.05$ & 49.72 & 49.76\\
 $50 \pm 0.10$ & 49.71& 49.74\\
 $50 \pm 0.15$ & 49.70& 49.72\\
 $50 \pm 0.20$ & 49.69 & 49.71\\
 $50 \pm 0.25$ & 49.68 & 49.69\\
 $50 \pm 0.30$ & 49.67 & 49.68 \\
 $50 \pm 0.35$ & 49.67 & 49.67\\
 \hline
\end{tabular}
\end{table}

\subsection{Formal Guarantees}

In order to compare the performance of our approach with the baseline controller in Fig.~\ref{fig:EVtikz}, we have simulated the GB model with baseline controller having different values for the deadband threshold. The steady-state values of frequency is reported in Table~\ref{table:2} for both unidirectional and bidirectional EVs. As can be seen, the highest steady-state frequency is achieved when deadband thresholds are both $50$ Hz, i.e., no deadband component in the baseline controller which requires instantaneous response from the EVs. Even in such a case, the baseline controller is unable to satisfy the specification $\psi$ in \eqref{eq:spec} as the steady-state is outside of the smaller target region $I_2$. 
In contrast, the multi-phase Controller based on the two symbolic controllers C1 and C2 satisfies the required specification. This comes at the cost of higher participation in comparison with the baseline controller as reported in Fig.~\ref{fig:bi_input} and Fig.~\ref{fig:uni_input}. 

\subsection{Robustness of the Controller}
\begin{figure}
\includegraphics[width=\columnwidth]{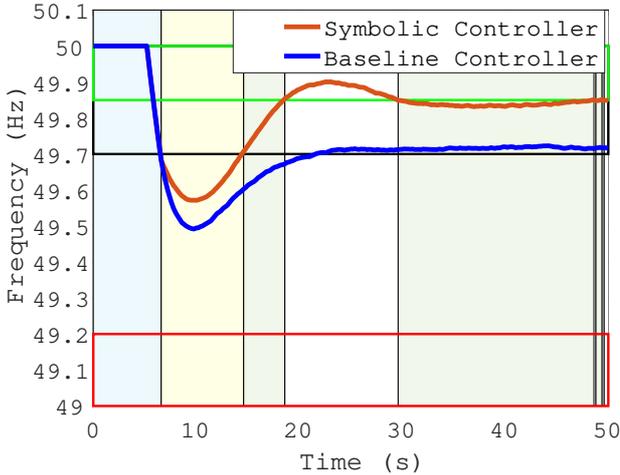}
\caption{Symbolic control for frequency regulation with bidirectionally charged EVs with up to $10\%$ uniformly distributed random uncertainty in participation. The frequency under the synthesised symbolic controller still satisfies the specification $\psi$ in \eqref{eq:spec} but the baseline controller fails to do so.}
\label{fig:uncertain_bi_freq}
\end{figure}

\begin{figure}
\includegraphics[width=\columnwidth]{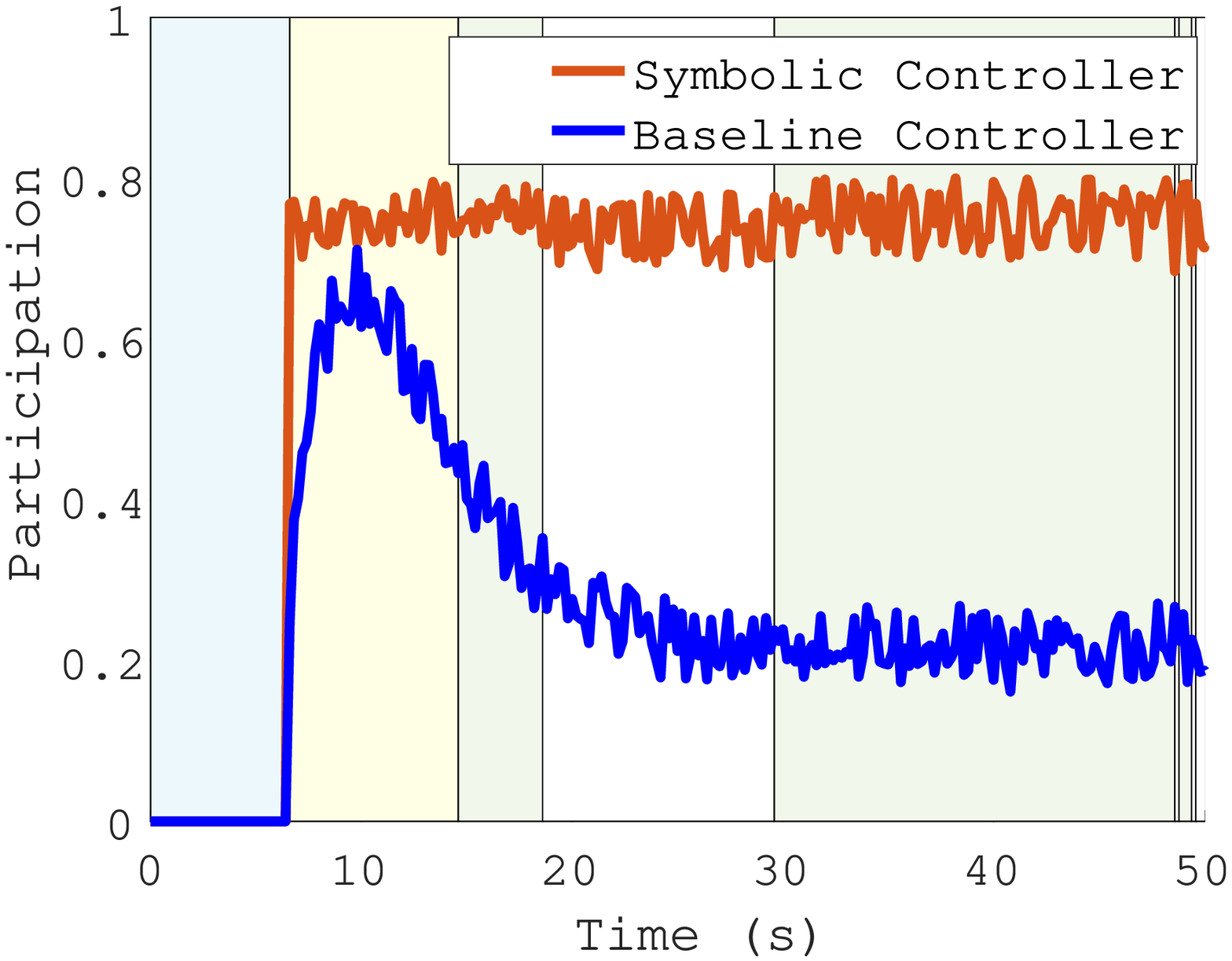}
\caption{Percentage of participation of bidirectionally charged EVs that has up to $10\%$ uniformly distributed random uncertainty in participation, as a function of time obtained from our synthesis approach to satisfy $\psi$ in \eqref{eq:spec} and from the baseline controller of Fig.~\ref{fig:EVtikz}.}
\label{fig:uncertain_bi_input}
\end{figure}

\begin{figure}
\includegraphics[width=\columnwidth]{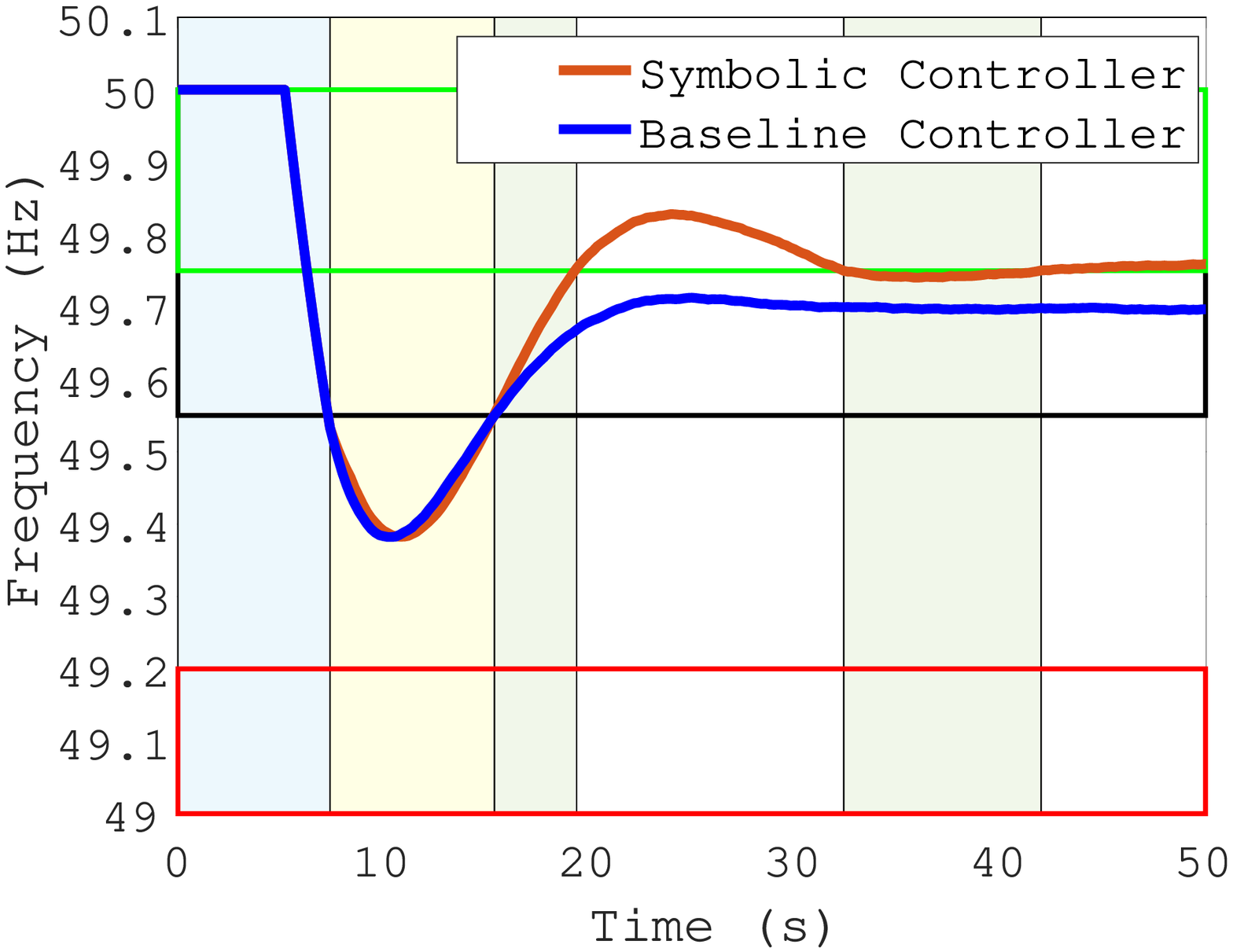}
\caption{
Symbolic control for frequency regulation with unidirectionally charged EVs that has up to $10\%$ uniformly distributed random uncertainty in participation. The frequency under the synthesised symbolic controller still satisfies the specification but the baseline controller fails to do so.}
\label{fig:uncertain_uni_freq}
\end{figure}

\begin{figure}
\includegraphics[width=\columnwidth]{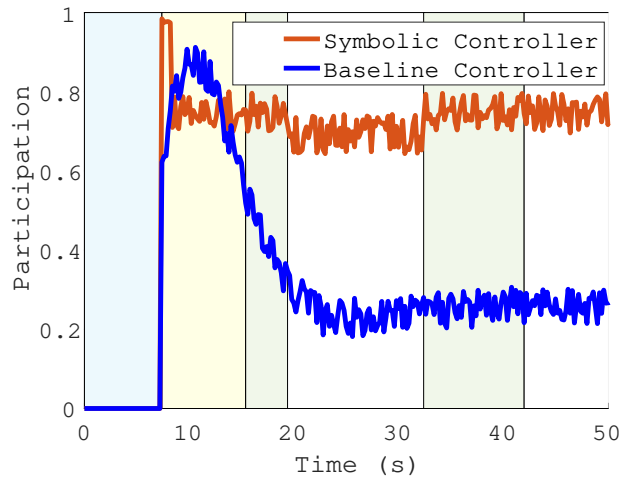}
\caption{Percentage of participation of unidirectionally charged EVs that has up to $10\%$ uniformly distributed random uncertainty in participation, as a function of time obtained from our synthesis approach to satisfy $\psi$ in \eqref{eq:spec} and from the baseline controller of Fig.~\ref{fig:EVtikz}.}
\label{fig:uncertain_uni_input}
\end{figure}

To measure the robustness of the controller against uncertainty in the participation of the EVs, we allow up to $10\%$ uncertainty in the participation required by the symbolic controller. Fig.~\ref{fig:uncertain_bi_freq} and Fig.~\ref{fig:uncertain_uni_freq} show that both cases of unidirectionally and bidirectionally charged EVs continue to facilitate satisfaction of the specification $\psi$, despite the uncertainty on the participation, although for bidirectional charging the specification is satisfied after a relatively longer time period ($\approx 48$ seconds). Fig.~\ref{fig:uncertain_bi_input} and Fig.~\ref{fig:uncertain_uni_input} show the variation in participation is substantial and that the fixed value assigned inside the winning region, can also fluctuate. Uncertainty has a larger effect on bidirectional charging than unidirectional charging as each bidirectional vehicle contributes double the power of its unidirectional equivalent. With increased uncertainty, the time taken to converge to the winning region also increases.

\medskip

Overall, our design approach encourages more refined specifications for frequency of the grid. It allows designing controllers automatically to satisfy those specifications with correctness guarantees and is more robust. Other approaches are unable to provide controllers automatically with correctness guarantees and require manual tuning of parameters while relying on simulations.


\section{Conclusion}
\label{sec:Conclusion}
In conclusion, our paper proposes a new approach for integration of EVs in frequency response services with the following features:
\begin{itemize}
    \item A proof of concept for the design and use of symbolic controllers in primary frequency response services;
    \item Using temporal logic to encode the requirements on the frequency that are usually expressed in natural language;
    \item Formal guarantees on satisfaction of such requirements under the synthesised symbolic controller;
    \item Enhanced response to large frequency loss events with symbolic control due to a more robust controller design;
    \item The controller is robustness against uncertainty in the EVs participation;
    \item Simulation results that show correctness of the controller design against a more refined specification on the GB grid.
\end{itemize}

In future work, we plan to conduct a quantitative analysis of uncertainty in satisfying the desired properties on the frequency. We also intend to extend the concepts used in this paper on nonlinear models of the power systems, real-time computation of the controllers, and study security issues associated with the implementation of these control schemes.

\section{Acknowledgement}
\label{sec:Acknowledgement}
The authors want to acknowledge Kaushik Mallik from Max Planck Institute and Mahmoud Khaled from TU Munich for their assistance with the use of SCOTS.
\bibliographystyle{ieeetr}
\bibliography{references}

\begin{thebibliography}{10}

\bibitem{MU201306}
Y.~{Mu}, J.~{Wu}, J.~{Ekanayake}, N.~{Jenkins}, and H.~{Jia}, ``Primary
  frequency response from electric vehicles in the {Great Britain} power
  system,'' {\em IEEE Transactions on Smart Grid}, vol.~4, pp.~1142--1150, June
  2013.

\bibitem{MUCHAOXU2018}
C.~{Mu}, W.~{Liu}, and W.~{Xu}, ``Hierarchically adaptive frequency control for
  an ev-integrated smart grid with renewable energy,'' {\em IEEE Transactions
  on Industrial Informatics}, vol.~14, pp.~4254--4263, Sep. 2018.

\bibitem{Izadkhast2015}
S.~{Izadkhast}, P.~{Garcia-Gonzalez}, and P.~{Frías}, ``An aggregate model of
  plug-in electric vehicles for primary frequency control,'' {\em IEEE
  Transactions on Power Systems}, vol.~30, pp.~1475--1482, May 2015.

\bibitem{EnergyEmergenciesExecutiveCommittee2020}
{Energy Emergencies Executive Committee}, ``{GB} power system disruption –
  {09 AUGUST 2019},'' Tech. Rep. January, Department for Business, Energy and
  Industrial Strategy, 2020.

\bibitem{CALLOWAY2011}
D.~S. {Callaway} and I.~A. {Hiskens}, ``Achieving controllability of electric
  loads,'' {\em Proceedings of the IEEE}, vol.~99, pp.~184--199, Jan 2011.

\bibitem{TAN201403}
Z.~{Tan}, P.~{Yang}, and A.~{Nehorai}, ``An optimal and distributed demand
  response strategy with electric vehicles in the smart grid,'' {\em IEEE
  Transactions on Smart Grid}, vol.~5, pp.~861--869, March 2014.

\bibitem{ISLAM201911}
M.~M. Islam, X.~Zhong, Z.~Sun, H.~Xiong, and W.~Hu, ``Real-time frequency
  regulation using aggregated electric vehicles in smart grid,'' {\em Computers
  \& Industrial Engineering}, vol.~134, pp.~11 -- 26, 2019.

\bibitem{UDDIN2017710}
K.~Uddin, T.~Jackson, W.~D. Widanage, G.~Chouchelamane, P.~A. Jennings, and
  J.~Marco, ``On the possibility of extending the lifetime of lithium-ion
  batteries through optimal {V2G} facilitated by an integrated vehicle and
  smart-grid system,'' {\em Energy}, vol.~133, pp.~710 -- 722, 2017.

\bibitem{KEMPTON2005268}
W.~Kempton and J.~Tomić, ``Vehicle-to-grid power fundamentals: Calculating
  capacity and net revenue,'' {\em Journal of Power Sources}, vol.~144, no.~1,
  pp.~268 -- 279, 2005.

\bibitem{Yilmaz2013}
M.~{Yilmaz} and P.~T. {Krein}, ``Review of battery charger topologies, charging
  power levels, and infrastructure for plug-in electric and hybrid vehicles,''
  {\em IEEE Transactions on Power Electronics}, vol.~28, pp.~2151--2169, May
  2013.

\bibitem{Bao2012}
K.~{Bao}, S.~{Li}, and H.~{Zheng}, ``Battery charge and discharge control for
  energy management in {EV} and utility integration,'' in {\em 2012 IEEE Power
  and Energy Society General Meeting}, pp.~1--8, July 2012.

\bibitem{BK08}
C.~Baier and J.-P. Katoen, {\em Principles of Model Checking}.
\newblock MIT Press, 2008.

\bibitem{Alur2000}
R.~{Alur}, T.~A. {Henzinger}, G.~{Lafferriere}, and G.~J. {Pappas}, ``Discrete
  abstractions of hybrid systems,'' {\em Proceedings of the IEEE}, vol.~88,
  no.~7, pp.~971--984, 2000.

\bibitem{Girard2007pappas}
A.~{Girard} and G.~J. {Pappas}, ``Approximation metrics for discrete and
  continuous systems,'' {\em IEEE Transactions on Automatic Control}, vol.~52,
  no.~5, pp.~782--798, 2007.

\bibitem{Mallik2019}
K.~{Mallik}, A.~{Schmuck}, S.~{Soudjani}, and R.~{Majumdar}, ``Compositional
  synthesis of finite-state abstractions,'' {\em IEEE Transactions on Automatic
  Control}, vol.~64, no.~6, pp.~2629--2636, 2019.

\bibitem{Ames_2019}
A.~D. Ames, S.~Coogan, M.~Egerstedt, G.~Notomista, K.~Sreenath, and P.~Tabuada,
  ``Control barrier functions: Theory and applications,'' {\em 2019 18th
  European Control Conference (ECC)}, Jun 2019.

\bibitem{lindemann2020control}
L.~Lindemann, G.~J. Pappas, and D.~V. Dimarogonas, ``Control barrier functions
  for nonholonomic systems under risk signal temporal logic specifications,''
  2020.

\bibitem{girard2005}
A.~Girard, ``Reachability of uncertain linear systems using zonotopes,'' in
  {\em Proceedings of the 8th International Conference on Hybrid Systems:
  Computation and Control}, HSCC’05, (Berlin, Heidelberg), p.~291–305,
  Springer-Verlag, 2005.

\bibitem{HS_TAC19}
S.~Haesaert and S.~Soudjani, ``Robust dynamic programming for temporal logic
  control of stochastic systems,'' {\em CoRR}, vol.~abs/1811.11445, 2018.

\bibitem{SSoudjani}
S.~{Soudjani}, {\em Formal Abstractions for Automated Verification and
  Synthesis of Stochastic Systems}.
\newblock PhD thesis, Technische Universiteit Delft, The Netherlands, 2014.

\bibitem{majumdar2019symbolic}
R.~Majumdar, K.~Mallik, and S.~Soudjani, ``Symbolic controller synthesis for
  {B}\"{u}chi specifications on stochastic systems,'' 2019.

\bibitem{lavaei2020formal}
A.~Lavaei, F.~Somenzi, S.~Soudjani, A.~Trivedi, and M.~Zamani, ``Formal
  controller synthesis for continuous-space {MDP}s via model-free reinforcement
  learning,'' {\em arXiv:2003.00712}, 2020.

\bibitem{kazemi2020formal}
M.~Kazemi and S.~Soudjani, ``Formal policy synthesis for continuous-space
  systems via reinforcement learning,'' {\em arXiv:2005.01319}, 2020.

\bibitem{Rungger2016}
M.~Rungger and M.~Zamani, ``{SCOTS}: A tool for the synthesis of symbolic
  controllers,'' in {\em Proceedings of the 19th International Conference on
  Hybrid Systems: Computation and Control}, HSCC ’16, (New York, NY, USA),
  p.~99–104, Association for Computing Machinery, 2016.

\bibitem{Kochdumper2020}
N.~Kochdumper, B.~Sch\"{u}rmann, and M.~Althoff, ``Utilizing dependencies to
  obtain subsets of reachable sets,'' in {\em Proceedings of the 23rd
  International Conference on Hybrid Systems: Computation and Control}, HSCC
  ’20, (New York, NY, USA), Association for Computing Machinery, 2020.

\bibitem{Roy2011}
P.~Roy, P.~Tabuada, and R.~Majumdar, ``Pessoa 2.0: A controller synthesis tool
  for cyber-physical systems,'' in {\em Proceedings of the 14th International
  Conference on Hybrid Systems: Computation and Control}, HSCC ’11, (New
  York, NY, USA), p.~315–316, Association for Computing Machinery, 2011.

\bibitem{Frehse2011}
G.~Frehse, C.~Le~Guernic, A.~Donz{\'e}, S.~Cotton, R.~Ray, O.~Lebeltel,
  R.~Ripado, A.~Girard, T.~Dang, and O.~Maler, ``{SpaceEx: Scalable
  Verification of Hybrid Systems},'' in {\em Computer Aided Verification}
  (G.~Gopalakrishnan and S.~Qadeer, eds.), (Berlin, Heidelberg), pp.~379--395,
  Springer Berlin Heidelberg, 2011.

\bibitem{FAUST15}
S.~Soudjani, C.~Gevaerts, and A.~Abate, ``\textsf{FAUST}$^{\textsf{2}}$: Formal
  abstractions of uncountable-state stochastic processes,'' in {\em TACAS'15},
  vol.~9035 of {\em Lecture Notes in Computer Science}, pp.~272--286, Springer,
  2015.

\bibitem{CauchiTACAS19}
N.~Cauchi and A.~Abate, ``{StocHy}: automated verification and synthesis of
  stochastic processes,'' in {\em 25th International Conference on Tools and
  Algorithms for the Construction and Analysis of Systems (TACAS)}, 2019.

\bibitem{PRISM}
M.~Kwiatkowska, G.~Norman, and D.~Parker, ``{PRISM} 4.0: Verification of
  probabilistic real-time systems,'' in {\em Proc. 23\textsuperscript{rd}
  International Conference on Computer Aided Verification (CAV'11)}
  (G.~Gopalakrishnan and S.~Qadeer, eds.), vol.~6806 of {\em LNCS},
  pp.~585--591, Springer, 2011.

\bibitem{ProbReach}
F.~Shmarov and P.~Zuliani, ``Probreach: Verified probabilistic
  delta-reachability for stochastic hybrid systems,'' in {\em Proceedings of
  the 18th International Conference on Hybrid Systems: Computation and
  Control}, (New York, NY, USA), p.~134–139, Association for Computing
  Machinery, 2015.

\bibitem{AMYTISS}
A.~Lavaei, M.~Khaled, S.~Soudjani, and M.~Zamani, ``{AMYTISS}: {P}arallelized
  automated controller synthesis for large-scale stochastic systems,'' in {\em
  Proc. 32nd International Conference on Computer Aided Verification (CAV)},
  LNCS, Springer, 2020.

\bibitem{Zonetti2019}
D.~{Zonetti}, A.~{Saoud}, A.~{Girard}, and L.~{Fribourg}, ``A symbolic approach
  to voltage stability and power sharing in time-varying dc microgrids,'' in
  {\em 2019 18th European Control Conference (ECC)}, pp.~903--909, 2019.

\bibitem{SA14TCL}
S.~Soudjani and A.~Abate, ``Aggregation and control of populations of
  thermostatically controlled loads by formal abstractions,'' {\em IEEE
  Transactions on Control Systems Technology}, vol.~23, no.~3, pp.~975--990,
  2015.

\bibitem{MESG13}
M.~Kamgarpour, C.~Ellen, S.~Soudjani, S.~Gerwinn, J.~Mathieu, N.~Mullner,
  A.~Abate, D.~Callaway, M.~Fr\"anzle, and J.~Lygeros, ``Modeling options for
  demand side participation of thermostatically controlled loads,'' in {\em
  International Conference on Bulk Power System Dynamics and Control (IREP)},
  pp.~1--15, August 2013.

\bibitem{Hackenberg2012}
G.~{Hackenberg}, M.~{Irlbeck}, V.~{Koutsoumpas}, and D.~{Bytschkow}, ``Applying
  formal software engineering techniques to smart grids,'' in {\em 2012 First
  International Workshop on Software Engineering Challenges for the Smart Grid
  ({SE-SmartGrids})}, pp.~50--56, June 2012.

\bibitem{Koutsoumpas2013}
V.~{Koutsoumpas} and P.~K. {Gupta}, ``Towards a constraint based approach for
  self-healing smart grids,'' in {\em 2013 2nd International Workshop on
  Software Engineering Challenges for the Smart Grid (SE4SG)}, pp.~17--24, May
  2013.

\bibitem{Akram2018}
W.~Akram and M.~A. Niazi, ``A formal specification framework for smart grid
  components,'' {\em Complex Adaptive Systems Modeling}, vol.~6, p.~5, Sep
  2018.

\bibitem{Uddin2019Riaz}
R.~Uddin, S.~A. Naseem, and Z.~Iqbal, ``Formal reliability analyses of power
  line communication network-based control in smart grid,'' {\em International
  Journal of Control, Automation and Systems}, vol.~17, pp.~3047--3057, Dec
  2019.

\bibitem{TheGridCode2020}
{National Grid}, ``{The Grid Code},'' Tech. Rep.~5, National Grid, 2020.

\bibitem{NationalGrid2016}
{National Grid}, ``{Future Requirements for Balancing Services},'' tech. rep.,
  National Grid, 2016.

\bibitem{statutoryinstrumentsno2665}
{National Grid}, ``The electricity safety, quality and continuity
  regulations,'' tech. rep., legislation.gov.uk, 2002.

\bibitem{NETS2019}
NETS, ``{National Electricity Transmission System Security and Quality of
  Supply Standard Version 2.4},'' tech. rep., National Grid, 1 April 2019.

\bibitem{Smith2016}
R.~Smith, ``{System Operability Framework 2016},'' Tech. Rep. November,
  National Grid, 2016.

\bibitem{kundur1994}
P.~Kundur, N.~Balu, and M.~Lauby, {\em Power System Stability and Control}.
\newblock EPRI power system engineering series, McGraw-Hill Education, 1994.

\bibitem{tabuada09}
P.~Tabuada, {\em Verification and Control of Hybrid Systems: A Symbolic
  Approach}.
\newblock Springer, 2009.

\end{thebibliography}

\appendix

\subsection{Linear Temporal Logic}
Consider a finite set of alphabets $\alphabeth$. Each alphabet evaluates a subset of states as true. Composed as an infinite string, these alphabets form infinite words defined as
$\word=\wordt{0},\wordt{1},\wordt{2},\ldots\in\alphabeth^{\mathbb{N}}$.
These words are connected to paths of system $S$ via a measurable labelling function $\Lab:X\rightarrow \alphabeth$ that assigns an alphabet $\alpha =\Lab(x)$ to state $x\in X$. That is, infinite paths $w = (x_0,u_0,x_1,u_1,\ldots)$ are mapped to infinite words $\alphabeth^{\mathbb N}$, as
$\word=\Lab(w) := (\Lab(x_0),\Lab(x_1),\Lab(x_2),\ldots)$.

\begin{definition}
	\label{def:LTL}
	An LTL formula over an alphabet $\alphabeth$ is constructed inductively as
	\begin{equation*}
	\label{eq:LTL}
	\psi ::=  \textsf{true} \,|\, p \,|\,\psi_1 \wedge \psi_2 \,|\, \neg \psi \,|\,  \nex \psi \,|\, \psi_1\until \psi_2,\quad p\in \Sigma,
	\end{equation*}
	with $\psi_1,\psi_2,\psi$ being LTL formulas.
	\end{definition}
\smallskip%

Let ${\word}_n=(\wordt{n},\wordt{n+1},\wordt{n+2},\ldots)$ be a postfix of the word $\word $. The satisfaction relation between $\word$ and a property $\psi$ is denoted by $ \word \vDash\psi$
(or equivalently $\word_0\vDash\psi$).
The semantics of the satisfaction relation $\vDash$ are defined recursively over $\word_n$ as follows
\begin{itemize}
\item $\word_n\vDash\textsf{true}$ always hold.
\item An atomic proposition, $ \word_n\vDash   p$  for $ p\in \Sigma$ holds if $p \in\wordt{n}$.
\item A logical conjunction, $\word_n\vDash \psi_1\wedge\psi_2$ holds
if $ \word_n\vDash \psi_1$ and $ \word_n\vDash \psi_2$.
\item A negation, $\word_n\vDash\neg p$ holds if $ \word_n\nvDash p$.
\item A temporal next operator, $\word_n\vDash\nex\psi $ holds if  
  $ \word_{n+1}\vDash \psi$.
\item A temporal until operator, $\word_n\vDash \psi_1\until\psi_2$  holds if there exists an $i \in \mathbb{N}$ such that $\word_{n+i} \vDash \psi_2$, and for all
$j \in\mathbb{N}$, $0\leq j<i$, we have $\word_{n+j}\vDash \psi_1$.
\end{itemize}

In addition to the aforementioned operators, we can also use \emph{disjunction} $\vee$, \emph{eventually} $\lozenge$, and \emph{always} $\square$ operators as $ \psi_1\vee\psi_2 = \neg(\neg\psi_1\wedge\neg\psi_2)$, $\lozenge\psi:=\left(\textsf{true} \until \psi\right)$ and $\square\psi:=\neg\left(\lozenge\neg\psi\right)$.

\end{document}